\documentclass[lettersize,journal]{IEEEtran}
\usepackage{amsmath,amsfonts}
\usepackage{algorithmic}
\usepackage{algorithm}
\usepackage{array}
\usepackage{subfig}
\usepackage{textcomp}
\usepackage{stfloats}
\usepackage{url}
\usepackage{verbatim}
\usepackage{graphicx}
\usepackage{cite}
\usepackage{hyperref}
\usepackage{xcolor}
\usepackage{caption}
\usepackage{tabularray}
\usepackage{color}
\hypersetup{
  colorlinks=true,
  linkcolor=blue,
  citecolor=blue,
  urlcolor=blue
}

\begin{document}

\title{NN-Based Frequency Domain DPD for OFDM Massive MIMO Transmitters With Multiple States}

\author{Yundi Zhang, Yanshi Sun,~\IEEEmembership{Member,~IEEE}, and Li Chen,~\IEEEmembership{Senior Member,~IEEE}
\thanks{Yundi Zhang and Li Chen are with the CAS Key Laboratory of Wireless Optical Communication, University of Science and Technology of China (USTC), Hefei 230027, China (e-mail: zyd027570@mail.ustc.edu.cn; chenli87@ustc.edu.cn).

Yanshi Sun is with the School of Computer Science and Information Engineering, Hefei University of Technology, Hefei 230009, China (e-mail: sys@hfut.edu.cn).}}



\maketitle

\begin{abstract}
Frequency domain (FD)-digital predistortion (DPD) is a low-complexity DPD solution for massive multiple-input-multiple-output (MIMO) transmitters (TXs). 
In this letter, we extend FD-DPD to scenarios with multiple signal states (e.g., bandwidths and power levels). 
First, we propose a new neural network (NN)-based FD-DPD model, whose main idea is to use a hypernetwork (HN) to generate parameters for the output layer of the main NN based on the signal states. 
Then, we introduce how to effectively train the model with the help of time-domain (TD)-DPD. 
Experimental results show that the proposed model can achieve excellent performance, without requiring additional online training when signal states change. 

\end{abstract}


\vspace{-15pt}

\section{Introduction}\label{S1}
\IEEEPARstart{I}n radio frequency (RF) transmitters (TXs), the nonlinearity of power amplifier (PA) is a severe hardware impairment. 
Digital predistortion (DPD) is an effective method to compensate for the harmful effect of nonlinearities, which has attracted tremendous research interest. 
For example, authors in \cite{b1} and \cite{b2} proposed a memory polynomial (MP) model and a neural network (NN)-based model for DPD, respectively. 
Some further DPD studies\cite{b13}\cite{b14} considered scenarios with multiple signal states (e.g., bandwidths and power levels). 
But these DPD methods are all implemented in the time domain (TD), and each PA requires a separate DPD module. 
In massive multiple-input-multiple-output (MIMO) TXs, which typically contain hundreds of PAs, 
deploying an individual TD-DPD for each PA would incur substantial hardware and computational costs. 

To alleviate this issue, frequency-domain (FD)-DPD has recently been proposed, which jointly suppresses nonlinearities of all PAs in FD.  
Researchers in \cite{b6} proposed an MP-based FD-DPD scheme for 
MIMO TXs with single user. 
In \cite{b5}, authors proposed an NN-based FD-DPD model, which is referred to as FD-NN. 
The authors in \cite{b3} proposed an FD-DPD model based on convolutional neural networks (CNN), and they extended their work in \cite{b4}. 
However, these works didn't take into account the dynamic changes in signal states, which are common in modern communication systems, 
and the performance of these proposed methods degrades after the signal states change. 
On the other hand, in some studies on TD-DPD\cite{b14}, when encountering a change in signal states, the authors attempted to update the DPD model by feeding back the PA's output signal, 
and it is difficult to adopt such a strategy in FD-DPD because the TD signal being fed back cannot be directly used to train the FD-DPD model. 
Therefore, an FD-DPD model that can be directly applied to multiple signal states after offline training, without online updates, is crucial, which motivates our work. 

In this letter, we propose a hypernetwork (HN)-based FD-NN model, which can be used to perform FD-DPD for fully digital orthogonal frequency-division multiplexing (OFDM)-based MIMO TXs with multiple signal states. 
An efficient method to train the proposed model is also provided. 
Experimental results show that, whether in single-user or multi-user scenarios, 
the proposed model maintains excellent performance across varying signal states without the need for retraining. 

\vspace{-10pt}

\section{System Model}\label{S2}
\subsection{MU-MIMO-OFDM System Model}\label{S2-A}

\begin{figure}[t]
    \centering
    \subfloat[No DPD with ideal PAs.]{\label{fig_system_model_a}
    \includegraphics[width=0.90\linewidth]{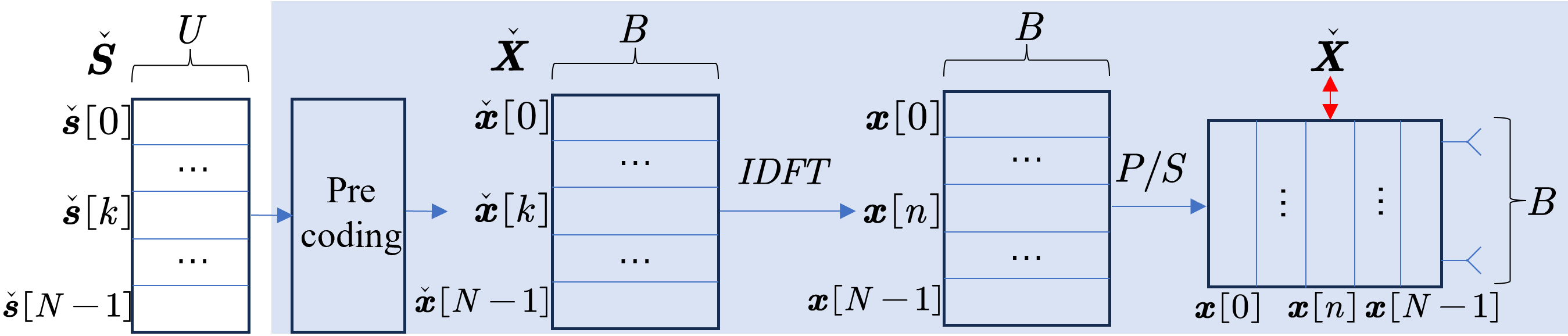}}

    \vspace{-9pt}

    \subfloat[FD-DPD with nonlinear PAs.]{\label{fig_system_model_b}
    \includegraphics[width=0.98\linewidth]{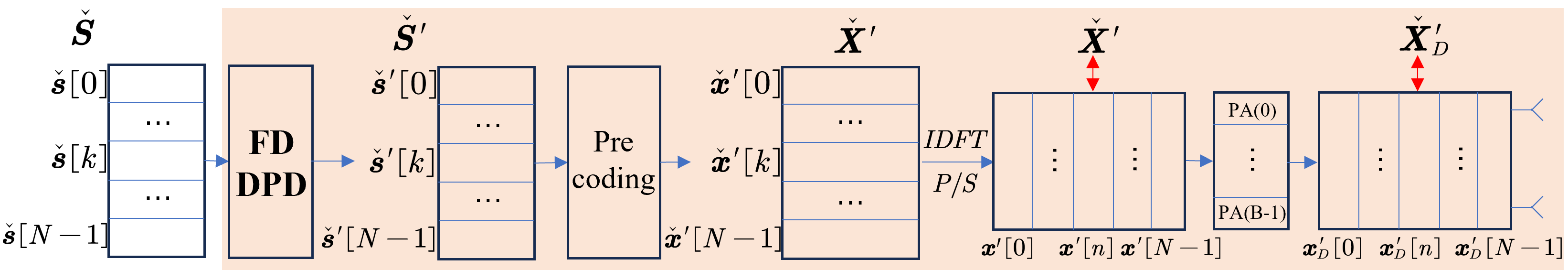}}

    \vspace{-9pt}

    \subfloat[TD-DPD with nonlinear PAs.]{\label{fig_system_model_c}
    \includegraphics[width=0.90\linewidth]{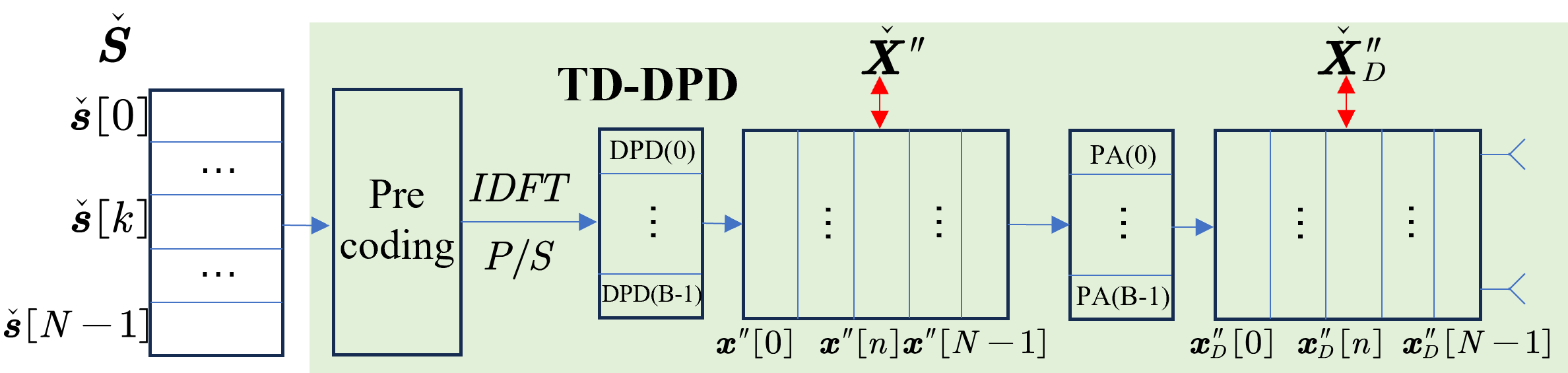}}

    \caption{System model of a fully digital MIMO TX. }
    \label{fig_system_model}
\end{figure}

As shown in Fig.~\ref{fig_system_model_a}, we first consider a fully digital MIMO TX with ideal PAs, which has $B$ antennas serving $U\!\ll\!B$ users simultaneously
\footnote{Without loss of generality, we restrict the presentation below to one OFDM symbol.}. 
Each OFDM symbol consists of $N$ subcarriers, among which $N_\text{d}$ are used as data subcarriers, $\boldsymbol{\check{s}}[k]\!\in\!\mathbb{C}^{U}$ represents the FD symbol vector of all users at the $k$-th subcarrier, 
and we denote $\boldsymbol{\check{S}}\!=\![ \boldsymbol{\check{s}}[ 0 ] ;\ldots;\boldsymbol{\check{s}}[ N-1 ] ]\in\mathbb{C}^{N\times U}$ as the FD symbol matrix 
\footnote{$\boldsymbol{\check{s}}[ k ]$ is viewed here as a row vector. In the remainder of this letter, one-dimensional vectors are sometimes treated as row vectors, which readers can easily determine from the context.}.

Then, digital precoding is applied for each FD symbol vector. Let $\boldsymbol{\check{W}}[ k ]\!\in\!\mathbb{C}^{B\times U}$ be the precoding matrix for subcarrier $k$, 
we have $\boldsymbol{\check{x}}[ k ]\!=\!\boldsymbol{\check{W}}[ k ] \boldsymbol{\check{s}}[ k ]\!\in\!\mathbb{C}^{B}$. 
$\boldsymbol{\check{W}}[ k ]$ can be obtained through different methods. In this letter, zero-forcing (ZF) precoding is considered, so $\boldsymbol{\check{W}}[ k ]$ can be expressed as
\begin{equation}\label{eq1}
    \boldsymbol{\check{W}}[ k ] =\alpha \boldsymbol{\check{H}}^\text{H}[ k ] ( \boldsymbol{\check{H}}[ k ] \boldsymbol{\check{H}}^\text{H}[ k ] ) ^{-1}, 
\end{equation}
where $\boldsymbol{\check{H}}[ k ]\!\in\!\mathbb{C}^{U\times B}$ is the FD channel matrix\cite{b4} for all users at subcarrier $k$, $\alpha$ is the normalization factor to meet the power constraint of the TX. 

After precoding, $N$-size inverse discrete Fourier transform (IDFT) and parallel-to-serial (P/S) conversion is applied 
to obtain a set of TD symbol vectors, i.e., $\{ \boldsymbol{x}[ n ] \} _{n=0}^{N-1}$, 
where $\boldsymbol{x}[ n ]\!\in\!\mathbb{C}^B$ is the TD symbol vector at time sample $n$. 
For the convenience of subsequent discussion, the result of performing S/P conversion and DFT on a set of TD symbol vectors is referred to as the {\it{FD equivalent form}}. 
For example, the FD equivalent form of $\{ \boldsymbol{x}[ n ] \} _{n=0}^{N-1}$ is obviously $\boldsymbol{\check{X}}\!=\![ \boldsymbol{\check{x}}[ 0 ] ;\ldots;\boldsymbol{\check{x}}[ N-1 ] ]\!\in\!\mathbb{C}^{N\times B}$. 

\vspace{-8pt}

\subsection{FD-DPD With Fixed State}\label{S2-B}
As shown in Fig.~\ref{fig_system_model_b}, we next consider the case where PA nonlinearity exists and FD-DPD is used to mitigate this problem. 
$\boldsymbol{\check{S}}^{\mkern1mu\prime}=[ \boldsymbol{\check{s}}^{\mkern1mu\prime}[ 0 ] ;\ldots;\boldsymbol{\check{s}}^{\mkern1mu\prime}[ N-1 ] ] \in \mathbb{C}^{N\times U}$ 
is the FD-DPD's results, where $\boldsymbol{\check{s}}^{\mkern1mu\prime}[ k ] \in \mathbb{C}^U$ is the FD symbol vector of all users at the $k$-th subcarrier after FD-DPD. 
Subsequently, similar to the discussion in Section \ref{S2-A}, digital precoding, $N$-size IDFT and P/S conversion are performed sequentially. Then we obtain a set of TD symbol vectors, 
i.e., $\{ \boldsymbol{x}^{\mkern1mu\prime}[ n ] \} _{n=0}^{N-1}$, which is the input of the PAs, and 
$\{ \boldsymbol{x}^{\mkern1mu\prime}_{\text{D}}[ n ] \} _{n=0}^{N-1}$ is the output of the PAs. 

Let $\boldsymbol{\check{X}}^{\mkern1mu\prime}$ and $\boldsymbol{\check{X}}^{\mkern1mu\prime}_{\text{D}} \in \mathbb{C}^{N\times B}$ be the FD equivalent forms of $\{ \boldsymbol{x}^{\mkern1mu\prime}[ n ] \} _{n=0}^{N-1}$ 
and $\{ \boldsymbol{x}^{\mkern1mu\prime}_{\text{D}}[ n ] \} _{n=0}^{N-1}$, respectively, we can derive the following relationship, i.e., 
\begin{equation}\label{eq2}
    \begin{split}
        \boldsymbol{\check{X}}_{\text{D}}^{\mkern1mu\prime} & =\boldsymbol{f}_{\text{FD-PA}}( \boldsymbol{\check{X}}^{\mkern1mu\prime} )\\
                                    & =\boldsymbol{f}_{\text{FD-PA}}( \boldsymbol{f}_{\text{precod.}}( \boldsymbol{\check{S}}^{\mkern1mu\prime} ) )\\
                                    & =\boldsymbol{f}_{\text{FD-PA}}( \boldsymbol{f}_{\text{precod.}}( \boldsymbol{f}_{\text{FD-DPD}}( \boldsymbol{\check{S}}; \boldsymbol{\lambda} ) ) ), 
    \end{split}
\end{equation}
where $\boldsymbol{f}_{\text{FD-PA}}:\mathbb{C}^{N\times B}\rightarrow \mathbb{C}^{N\times B}$ represents the FD input-output relationship of the MIMO TX's PA array, 
while $\boldsymbol{f}_{\text{precod.}}:\mathbb{C}^{N\times U}\rightarrow \mathbb{C}^{N\times B}$ and 
$\boldsymbol{f}_{\text{FD-DPD}}:\mathbb{C}^{N\times U}\rightarrow \mathbb{C}^{N\times U}$ denote the input-output relationship of digital precoding and FD-DPD, respectively. 
$\boldsymbol{\lambda}$ is the FD-DPD model's parameters. 

From the transmitter's perspective, we want the effects of FD-DPD and the nonlinearity of PAs to mutually cancel out, leaving only the effect of digital precoding. 
Therefore, our goal is to train $\boldsymbol{f}_{\text{FD-DPD}}$ such that $\boldsymbol{\check{X}}_{\text{D}}^{\mkern1mu\prime}=\boldsymbol{\check{X}}$, i.e., 
\begin{equation}\label{eq3}
    \boldsymbol{f}_{\text{FD-PA}}( \boldsymbol{f}_{\text{precod.}}( \boldsymbol{f}_{\text{FD-DPD}}( \boldsymbol{\check{S}} ; \boldsymbol{\lambda}) ) ) =\boldsymbol{f}_{\text{precod.}}( \boldsymbol{\check{S}} ), 
\end{equation}
holds as much as possible. 

\vspace{-8pt}

\subsection{FD-DPD With Multiple States}\label{S2-C}
When signal states change, the TD input-output relationship of PAs will change. Therefore, their FD input-output relationship, 
i.e., $\boldsymbol{f}_{\text{FD-PA}}$, will also change. 
Assume that the superscript $i$ indicates one specific state, 
equation \eqref{eq3} can be rewritten as 
\begin{equation}\label{eq4}
    \boldsymbol{f}_{\text{FD-PA}}^{(i)}( \boldsymbol{f}_{\text{precod.}}( \boldsymbol{f}_{\text{FD-DPD}}( \boldsymbol{\check{S}}^{(i)};\boldsymbol{\lambda} ) ) )=\boldsymbol{f}_{\text{precod.}}( \boldsymbol{\check{S}}^{(i)} ).
\end{equation}
In order to achieve excellent calibration performance within all states, equation \eqref{eq4} must hold as much as possible for every value of $i$, 
and the characteristics of $\boldsymbol{f}_{\text{FD-PA}}^{(i)}$ and $\boldsymbol{f}_{\text{FD-DPD}}$ must match each other within all 
states, which is much more challenging than the case when signal states are fixed.

\section{Proposed Dynamic FD-DPD}\label{S3}
\subsection{The Proposed FD HN-R2TDNN}
As described in Section \ref{S2-C}, the goal is to make $\boldsymbol{f}_{\text{FD-DPD}}$ and $\boldsymbol{f}_{\text{FD-PA}}^{(i)}$ match each other within all signal states,  
so we use the signal states of the current OFDM symbol, 
i.e., 
\begin{equation}\label{eq5}
    \boldsymbol{c}=[ c_1,c_2 ] =\left[ \frac{\text{BW}\text{(MHz)}}{\text{BW}_{\text{max}}\text{(MHz)}},\frac{\text{P}\text{(mW)}}{\text{P}_{\text{max}}\text{(mW)}} \right],
\end{equation} 
as part of the input to $\boldsymbol{f}_{\text{FD-DPD}}$. 
Now, the FD-DPD model can be represented as follows: 
\begin{equation}\label{eq6}
    \boldsymbol{\check{S}}^{\mkern1mu\prime}=\boldsymbol{f}_{\text{FD-DPD}}( \boldsymbol{\check{S}},\boldsymbol{c};\boldsymbol{\lambda} ). 
\end{equation}

\begin{figure}[t]
    \centering
    \subfloat[Proposed HN FD-NN for FD-DPD with multiple states.]{\label{fig_NN}
    \includegraphics[width=0.98\linewidth]{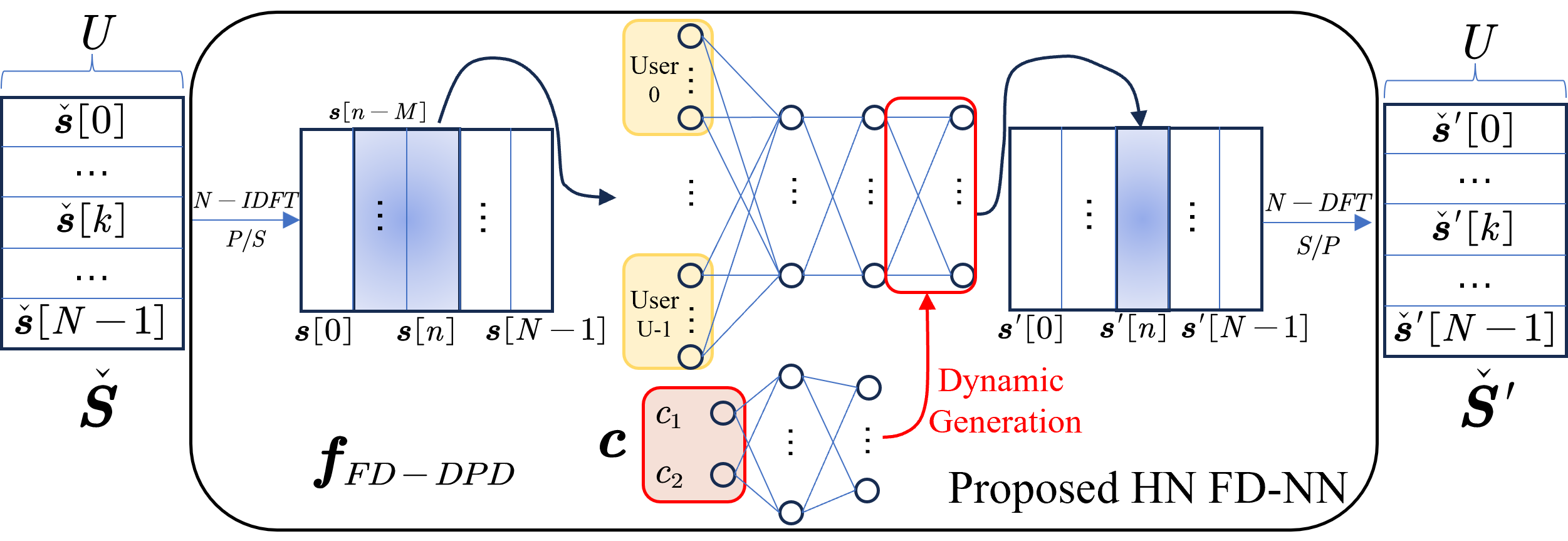}}

    \vspace{-10pt}

    \subfloat[Data loading method for training the HN FD-NN.]{\label{fig_loading}
    \includegraphics[width=0.8\linewidth]{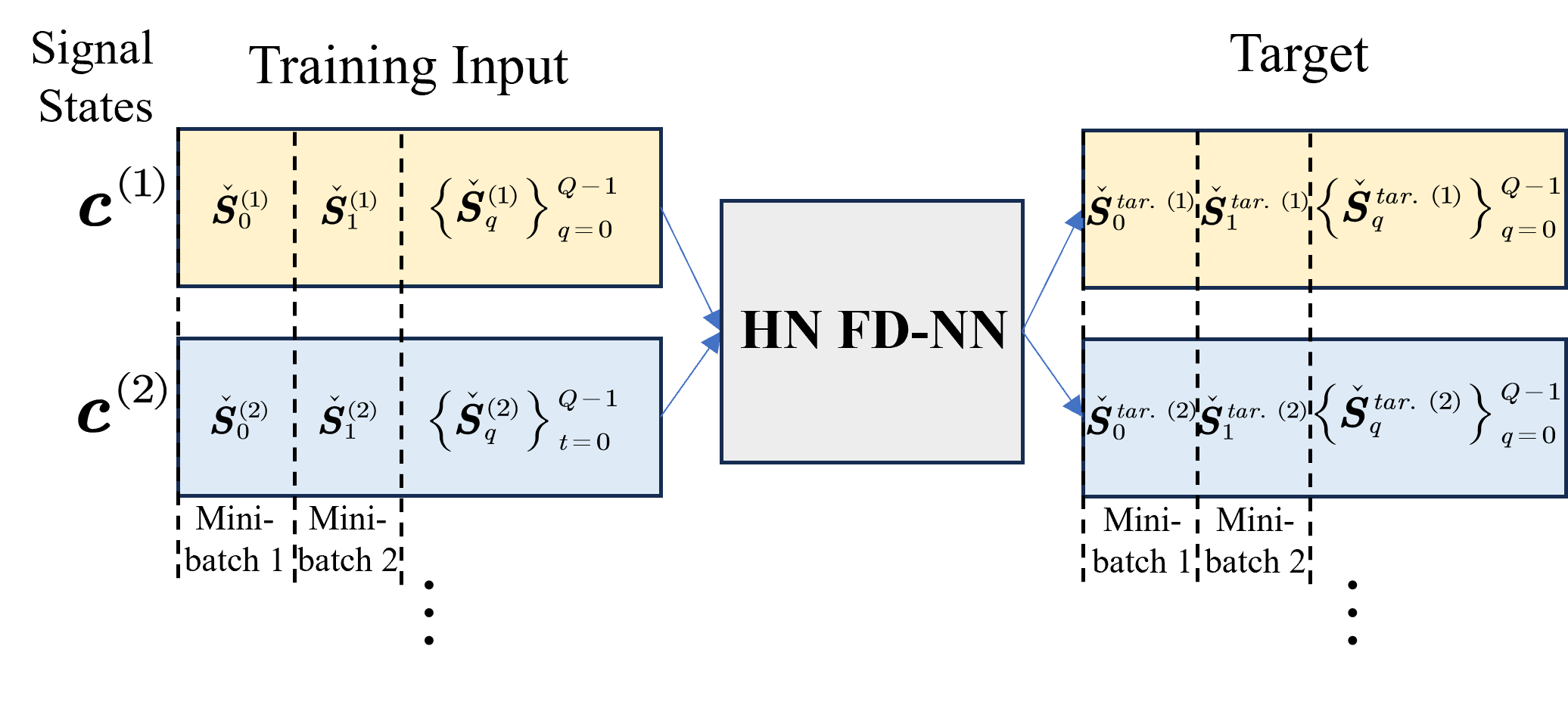}}

    \caption{Proposed method. }
    \label{fig_proposed}
\end{figure}

As shown in Fig.~\ref{fig_NN}, the details of the proposed HN FD-NN can be described as follows. 
First, we obtain the TD symbol vector set $\{ \boldsymbol{s}[ n ] \} _{n=0}^{N-1}$ 
by performing an $N$-size IDFT and P/S conversion on the FD symbol matrix $\boldsymbol{\check{S}}$. 

After that, the TD symbol vector with memory undergoes a real-valued decomposition. Specifically, we have 
\begin{equation}\label{eq7}
    \boldsymbol{z}_1[ n ]=[ \boldsymbol{s}_\text{R}[ n ],\ldots,\boldsymbol{s}_\text{R}[ n-M ],\boldsymbol{s}_\text{I}[ n ],\ldots,\boldsymbol{s}_\text{I}[ n-M ] ], 
\end{equation}
where $\boldsymbol{s}_\text{R}[ n ]$ and $\boldsymbol{s}_\text{I}[ n ] \in \mathbb{R}^{U}$ are the real and imaginary parts of $\boldsymbol{s}[ n ]$, 
and $M$ is the memory length of the whole model. 

Then $\boldsymbol{z}_1[ n ]$ is fed into an NN to obtain the predistorted TD vector $\boldsymbol{s}^{\mkern1mu\prime}[ n ]$. 
This NN is referred to as the {\it{main NN}}, whose output layer's weights and biases are generated by another NN called the {\it{hypernetwork (HN)}}, and the input of the HN is the signal state $\boldsymbol{c}$. 
All layers of the main NN and the HN are fully connected. 
Specifically, assume the main NN has a total of $G$ layers, and there are $G-2$ hidden layers, one input layer and one output layer. 
$D_{g}$ denotes the number of neurons in layer $g$, and $D_{1}=2(M+1)U$, $D_{G}=2U$. 
$\boldsymbol{W}_g\in \mathbb{R}^{D_g\times D_{g-1}}$ is the weight matrix that connects layer $g-1$ and $g$, 
$\boldsymbol{b}_g$ and $\boldsymbol{z}_g$ are the biases and outputs of layer $g$, respectively, 
and $2\leq  g\leq  G$. 
For $2\leq  g\leq  G-1$, we have 
\begin{equation}\label{eq9}
    \boldsymbol{z}_{g}=\phi ( \boldsymbol{W}_{g}\boldsymbol{z}_{g-1}+\boldsymbol{b}_{g} ),
\end{equation}
where $\phi(\cdot )$ is the activation function of the main NN's hidden layers. 
The output layer of the main NN uses a linear activation function to output a full range of values, 
so 
\begin{equation}
    \begin{split}
        \boldsymbol{z}_G=\boldsymbol{W}_G( \boldsymbol{c} ) \boldsymbol{z}_{G-1}+\boldsymbol{b}_G( \boldsymbol{c} )=[\boldsymbol{s}^{\mkern1mu\prime}_\text{R}[ n ],\boldsymbol{s}^{\mkern1mu\prime}_\text{I}[ n ]], 
    \end{split}
\end{equation}
where $\boldsymbol{W}_G( \boldsymbol{c} )$ and $\boldsymbol{b}_G( \boldsymbol{c} )$ are the weights and biases of the main NN's output layer, 
and they are also the HN's outputs, which can be dynamically changed according to the signal states $\boldsymbol{c}$. 
After obtaining $\{ \boldsymbol{s}^{\mkern1mu\prime}[ n ] \} _{n=0}^{N-1}$, performing $N$-size DFT and S/P conversion on it yields the output of the FD-DPD, i.e., $\boldsymbol{\check{S}}^{\mkern1mu\prime}$. 

Note that $\boldsymbol{W}_G$ and $\boldsymbol{b}_G$ are not trainable parameters, but rather the HN's inference results, which are involved in the output computation of the main NN, 
and the main NN and the HN can be trained as a whole. 
Moreover, although incorporating an HN increases the overall model complexity, 
the HN only requires to recalculate when signal states change. Therefore, for most of the time, 
the inference complexity of the entire model remains equivalent to that without the HN.

\vspace{-5pt}

\subsection{Offline Training Method}
First, we introduce how to obtain the FD-DPD's target output $\boldsymbol{\check{S}}^{\text{tar.}}$.
As shown in Fig.~\ref{fig_system_model_c}, now we assume that the TD-DPD model has already been trained within the current signal state, 
then we can obtain the TD-DPD's output symbol vectors, i.e., 
$\{ \boldsymbol{x}^{\mkern1mu\prime\prime}[ n ] \} _{n=0}^{N-1}$. 
We hope the performance of FD-DPD can catch up with that of TD-DPD,  
so the target value of $\{ \boldsymbol{x}^{\mkern1mu\prime}[ n ] \} _{n=0}^{N-1}$ can be $\{ \boldsymbol{x}^{\mkern1mu\prime\prime}[ n ] \} _{n=0}^{N-1}$. 
Therefore, we can derive $\boldsymbol{\check{S}}^{\text{tar.}}$ by applying S/P conversion, $N$-size DFT and the inverse transform of digital precoding to $\{ \boldsymbol{x}^{\mkern1mu\prime\prime}[ n ] \} _{n=0}^{N-1}$. 
Specifically, assuming that after S/P conversion and $N$-size DFT, the result on the $k$-th subcarrier is $\boldsymbol{\check{x}}^{\mkern1mu\prime\prime}[ k ]\in\mathbb{C}^B$. 
Then let $\boldsymbol{\check{s}}^{\text{tar.}}[ k ] =( \boldsymbol{\check{W}}[ k ] ) ^{\dag}\boldsymbol{\check{x}}^{\mkern1mu\prime\prime}[ k ]$, 
where $( \boldsymbol{\check{W}}[ k ] ) ^{\dag}$ is the psuedo-inverse of the precoding matrix $\boldsymbol{\check{W}}[ k ]$, 
and $\boldsymbol{\check{S}}^{\text{tar.}}=[ \boldsymbol{\check{s}}^{\text{tar.}}[ 0 ] ;\ldots;\boldsymbol{\check{s}}^{\text{tar.}}[ N-1 ] ]$. 

Next, we introduce how to train the proposed model. 
We first select several typical signal states as training states, then collect data for each state. 
As shown in Fig.~\ref{fig_loading}, 
let the training input of the model within the $i$-th training state be $\{ \boldsymbol{\check{S}}_{q}^{( i )} \} _{q=0}^{Q-1}$, 
where $q$ is the time index of the OFDM symbol, 
then we use the method mentioned above to obtain the corresponding target output $\{ \boldsymbol{\check{S}}_{q}^{\text{tar.}\ ( i )} \} _{q=0}^{Q-1}$, and 
the loss function for the $i$-th training state can be defined as  
\begin{equation}\label{eq10}
    L_i( \boldsymbol{\lambda } ) =\sum_{q=0}^{Q-1}{\lVert \boldsymbol{\check{S}}_{q}^{\text{tar.}\ ( i )}-\boldsymbol{f}_{\text{FD-DPD}}( \boldsymbol{\check{S}}_{q}^{( i )},\boldsymbol{c}^{( i )};\boldsymbol{\lambda } ) \rVert}_{\text{F}}^{2}, 
\end{equation}
where $\lVert \cdot \rVert _\text{F}$ represents the Frobenius norm. 
To prevent the model from gradually forgetting certain states during training, as shown in Fig.~\ref{fig_loading}, each minibatch loaded during training must contain data from all training states simultaneously. 
Thus, the overall loss function is 
\begin{equation}\label{eq11}
    L( \boldsymbol{\lambda } ) =\sum_i{L_i( \boldsymbol{\lambda } )}.
\end{equation}
Therefore, we can train the whole model by minimizing $L( \boldsymbol{\lambda } )$.

\vspace{-5pt}

\section{Numerical Results}\label{S4}
\subsection{Simulation Setup}
\begin{figure}[t]
    \centering
    \includegraphics[width=0.65\linewidth]{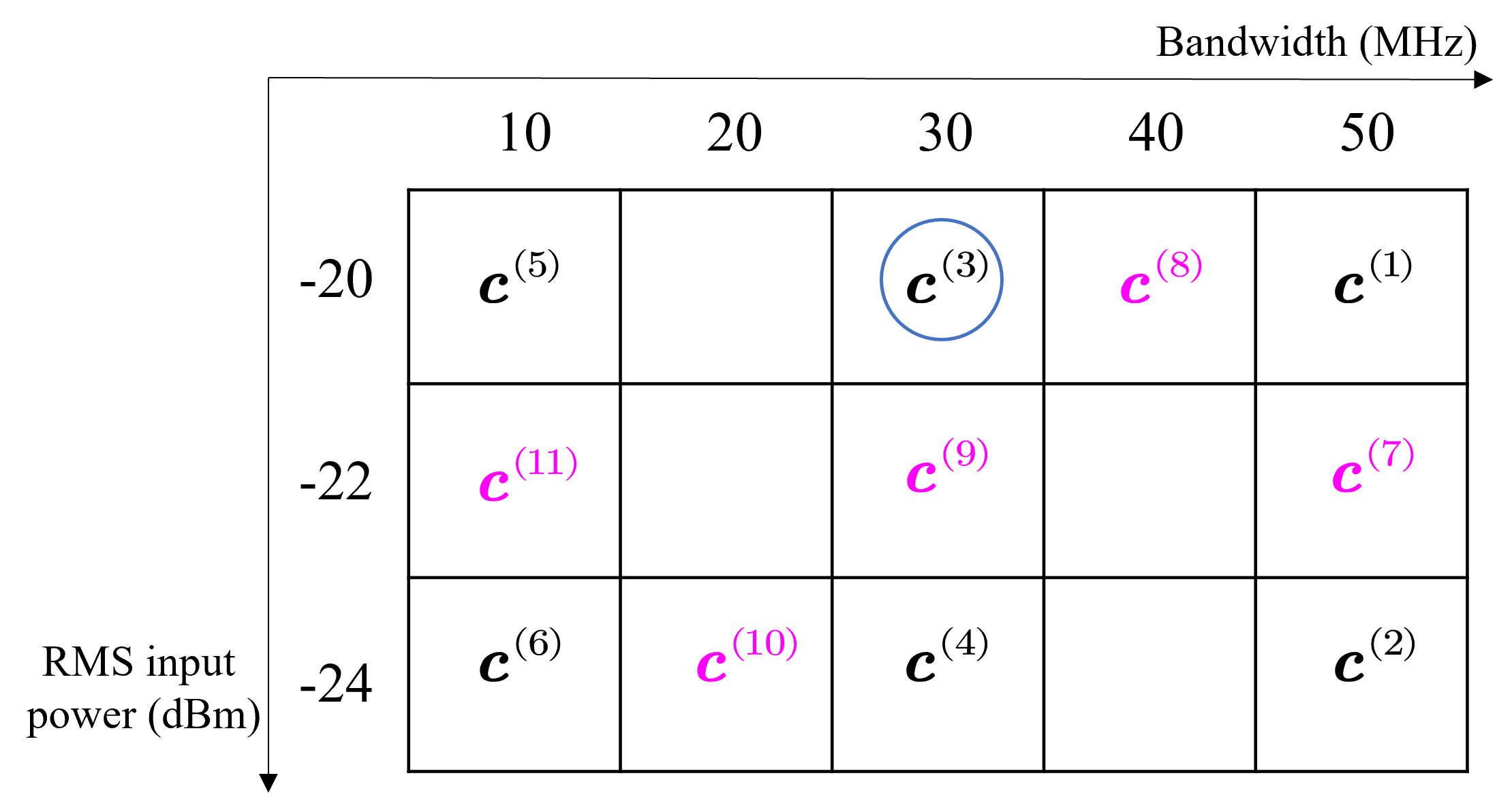}
    \caption{Signal states considered in this letter.} 
    \label{fig_signal_states}
\end{figure}
We consider 11 different signal states, as shown in Fig.~\ref{fig_signal_states}, and 
we consider a set of simulated PAs using the MP model\cite{b1}, whose coefficients are estimated using real measurements from the RF WebLab\cite{b9}. 
To accurately reflect the characteristics of PA under different states, we trained separate MP models for each signal state in Fig.~\ref{fig_signal_states} with different memory lengths and nonlinear orders. 
Specifically, 
For signal bandwidths of 10, 20, 30, 40, and 50 MHz, the MP model's memory lengths are set to 3, 4, 5, 6, and 7, respectively. The nonlinear orders are 7, 7, and 5 for RMS powers of -20, -22, and -24 dBm, respectively.

We train these MP models and validate the proposed method using MATLAB generated OFDM signals with a sampling rate of $200$ MHz and an IDFT size of $32,768$. 
The number of data subcarriers varies depending on the different signal bandwidth. For example, when the number of data subcarriers is $8192$, the corresponding bandwidth is $50$ MHz. 
The RMS input power is specified via the MATLAB API provided by the RF WebLab. 

We model a frequency-flat LOS channel using the channel model described in \cite{b4}. 
The carrier frequency is set to $30$ GHz. The median channel gain is $-61.9$ dB at 1m, and the pathloss exponent $\alpha=2.1$. 
The thermal noise variance is $-174$ dBm/Hz, and all the receivers' noise figure is $7$ dB. 

It is widely recognized that higher out-of-band emission power can be tolerated in MIMO systems\cite{b4}. Therefore, following some existing studies\cite{b3}, we mainly focus on the in-band linearization performance of FD-DPD, 
and we use the error vector magnitude (EVM)\cite{b12} of all users' received signal to quantify this performance. 
Furthermore, the performance of FD-DPD is evaluated from the transmitter's perspective using the normalized mean square error \cite{b13} of the transmitted signal (TX-NMSE) as an additional metric. 

We compare the proposed HN FD-NN with FD-NN, and 
we consider two cases: with ($U$, $B$) = ($1$, $100$) and ($4$, $100$). 
In the single-user case, the user is located 25 m from the TX at an angle of 70°. 
For the multi-user case, all users are located 25 m from the TX with equal power allocation, and their angles are 70°, 50°, 40°, and 20°, respectively. 
The number of neurons in each layer for all NNs used in both cases is shown in Table \ref{t1}, the activation functions of 
the main NN and the HN of the HN FD-NN are tanh and ReLU, respectively, while the activation function of the FD-NN is also tanh.   
We use the stochastic gradient descent (SGD) method with the Adam optimizer to train these NNs. 
The HN FD-NN is trained using mixed data of the 6 black states in Fig.~\ref{fig_signal_states}, the FD-NN is trained under the state circled in blue, and the pink states do not appear in any of the training stages. 
\begin{table}[t]
    \centering
    \caption{Number of neurons in each layer for all NNs}
    \begin{tabular}{|c|c|c|} 
    \hline
                & Case 1: U=1 B=100                                                         & Case 2: U=4 B=100                                                             \\ 
    \hline
    HN FD-NN & \begin{tabular}[c]{@{}c@{}}Main NN:~16-50-6-2\\HN: 2-40-24-14\end{tabular} & \begin{tabular}[c]{@{}c@{}}Main NN:~64-130-10-8\\HN: 2-160-100-88\end{tabular}  \\ 
    \hline
    FD-NN        & 16-50-6-2                                                                  & 64-130-10-8                                                                    \\
    \hline
    \end{tabular}
    \label{t1}
\end{table}

\vspace{-8pt}

\subsection{Results}
\begin{figure}[t]
    \centering
    \subfloat[Case 1 ($U=1$, $B=100$).]{\label{fig_EVM_U1}
    \includegraphics[width=0.58\linewidth]{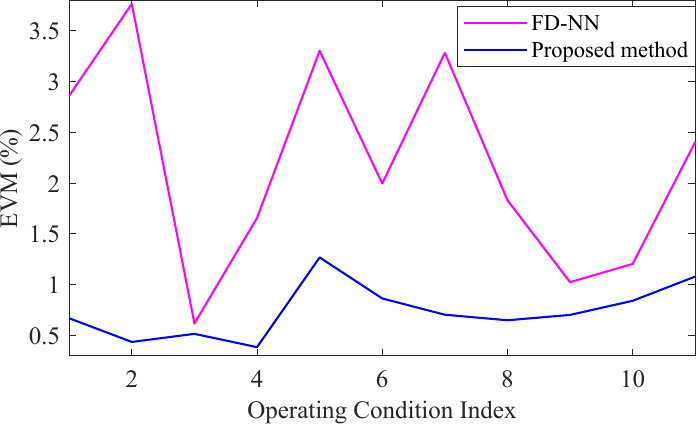}}

    \vspace{-5pt}

    \subfloat[Case 2 ($U=4$, $B=100$).]{\label{fig_EVM_U4}
    \includegraphics[width=0.58\linewidth]{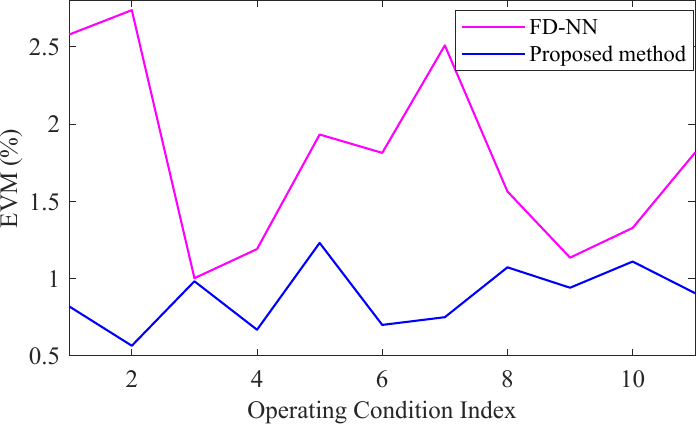}}

    \caption{The EVM results of different DPD methods. }
    \label{fig_EVM}
\end{figure}

\begin{figure}[t]
    \centering
    \includegraphics[width=0.58\linewidth]{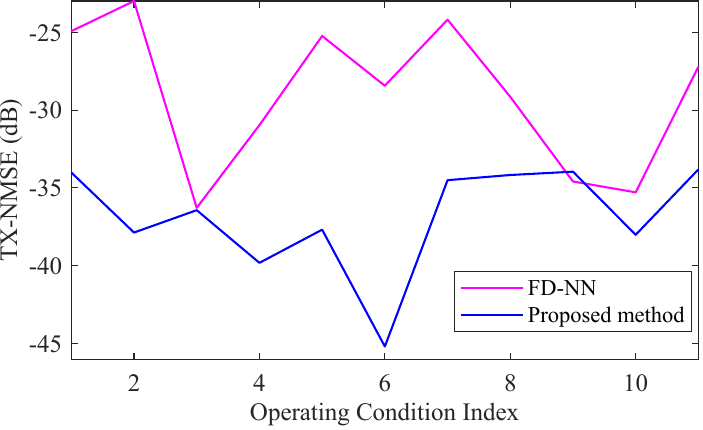}
    \caption{TX-NMSE results in Case 1.}
    \label{fig_Tx_NMSE}
\end{figure}

\begin{figure}[t]
    \centering
    \includegraphics[width=0.7\linewidth]{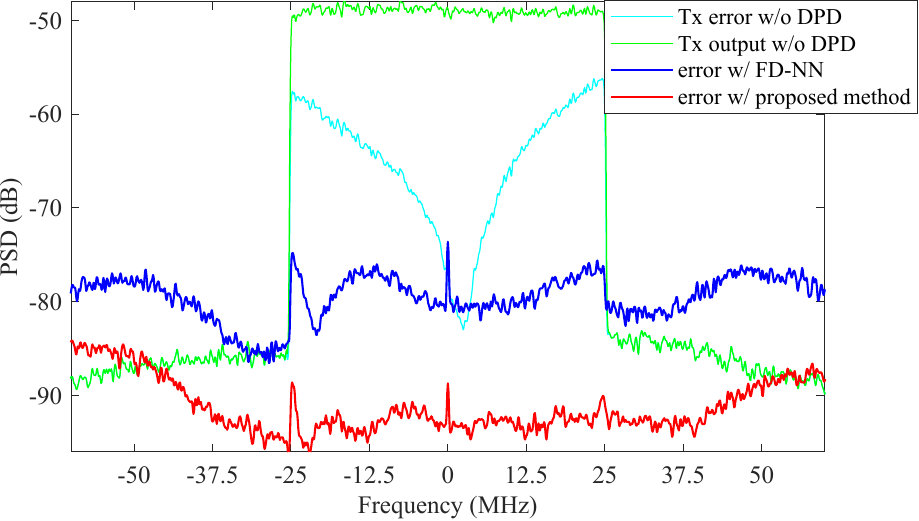}
    \caption{PSDs of the signals within state 1 ($50$ MHz and $-20$ dBm) in Case 1.}
    \label{fig_PSDs}
\end{figure}

Fig.~\ref{fig_EVM} shows the EVM of all users' received signal within all states. 
It can be seen that the proposed HN FD-NN achieves excellent performance within all states, 
even in those that did not appear during training (i.e., states 7 to 11). 
Whether the number of users is $1$ (single-user scenario) or $4$ (multi-user scenario), the proposed method can reduce the EVM to approximately $1$\% or even lower within all signal states. 
The TX-NMSE results, as shown in Fig.~\ref{fig_Tx_NMSE}, also leads to similar conclusions. 
The proposed method can reduce TX-NMSE to approximately $-35$ dB or lower within all signal states when the number of user is $1$. 
Fig.~\ref{fig_PSDs} shows the power spectral density (PSD) of the TX's output signal within state 1, and it also shows the PSDs of the error signals 
(obtained by subtracting the TX's ideal output signals from its actual output signals) for three scenarios: without DPD, with FD-NN, and with the proposed method. 
It can be seen that, within the relatively most challenging state 1 (which features the largest bandwidth and highest signal power), the proposed method can effectively reduce in-band error. 

Meanwhile, the FD-DPD with fixed state (i.e., FD-NN) performs well in its dedicated state (i.e., state 3) and certain similar states (e.g., state 9). 
However, its performance noticeably declines when encountering states significantly different from state 3 (e.g., state 1). 
As shown in Fig.~\ref{fig_PSDs}, within state 1, the FD-NN can only reduce in-band error to a certain extent. 
Furthermore, due to the mismatch between the FD-NN and the PA's characteristics within this state, the FD-NN may even slightly increase out-of-band emission power. 

\section{Conclusion}\label{S5}
In this letter, we have proposed a new multi-state FD-DPD model. 
The key idea is to utilize an HN to generate parameters for the output layer of the main NN based on signal states. 
With the help of TD-DPD's results, we can effectively train it offline. 
Numerical results have shown that the proposed method can reduce EVM to around 1\% or even lower within multiple signal states, whether in single-user or multi-user scenarios.

\end{document}